# Power Converter Topologies for Electrolyzer Applications to Enable Electric Grid Services

Bang L. H. Nguyen[†, §], Mayank Panwar[†], Rob Hovsapian[†], Kazunori Nagasawa[†], and Tuyen V. Vu[§]
[†]*National Renewable Energy Laboratory,* Golden, CO, USA
[§]*ECE Department, Clarkson University,* Potsdam, NY, USA
bang.nguyen@nrel.gov, mayank.panwar@nrel.gov, rob.hovsapian@nrel.gov,
kazunori.nagasawa@nrel.gov, tvu@clarkson.edu

*Abstract*—Hydrogen electrolyzers, with their operational flexibility, can be configured as smart dynamic loads which can provide grid services and facilitate the integration of more renewable energy sources into the electrical grid. However, to enable this ability, the electrolyzer system should be able to control both active and reactive power in coordination with the low-level controller of the electrolyzer via the power electronics system interface between the utility grid and electrolyzer. This paper discusses power converter topologies and the control scheme of this power electronics interface for electrolyzer applications to enable electricity grid services. For the sake of unity, in this paper, we consider the power converter system interfacing the utility grid at the line-to-line root mean square (RMS) value of 480 VAC–60 Hz and supplying to the 3500 A-750 kW PEM electrolyzer stack.

## I. Introduction

Hydrogen electrolysis is a mature technology that produces hydrogen from water using electricity. There are two main electrolyzer technologies, the alkaline and proton exchange membrane (PEM) electrolyzer. Although alkaline technology is mature with low manufacturing cost, it is slow at start-up and has problems of corrosion, and requires complicated maintenance, whereas PEM has a fast response and requires less maintenance [1]. Hydrogen itself can be used in various areas includes petroleum, ammonia, transportation, and power generation with a fuel cell or internal combustion engine. Additionally, hydrogen is considered an attractive and versatile form of energy storage owing to its low environmental impact [2]. Excess power can be converted and stored in form of hydrogen that can be used later in other industrial processes, transportation, and power generation. Hydrogen electrolyzers can be integrated with renewable and distributed energy resources (DER) to reduce the impact of variability on power quality and stability of the electric grid [3].

Furthermore, the electrolyzer can be operated as a smart dynamic load and provide grid services includes end-user energy management, transmission, and distribution system support, integration of renewables, and wholesale electric market services [4]. In [5], the authors propose a generic front-end controller for electrolyzer to enhance the grid flexibility. The work in [6] analyzes the 25 MW PEM electrolyzer installed in Belgium in providing grid services such as grid balance and frequency containment reserve. The potential of frequency support from an electrolyzer is investigated in [7] with a 1-MW pilot electrolyzer installed in the Netherlands. In [8], the hydrogen energy system is considered as a grid management tool to stabilize the grid frequency in the Hawaiian island grid. In [9], the integration between the distribution grid and electrolyzer fleet is employed to mitigate the impact of solar PV penetration by reducing overvoltage and voltage fluctuations.

To effectively serve the growing use and validation of electrolyzer applications, power electronics interfaces play a key role considering the control-ability and more efficient energy conversion. However, some of the existing literature has reviewed the power converter for electrolyzer itself [10]-[12] but the ability to support electricity grid services has not been considered. To fill the gap, this paper provides a discussion on the power converter topologies and control schemes for the power electronic interface for electrolyzer to enable grid services. The paper is organized as follows. Section II reviews the electrolyzer model and the baseline requirements of a power electronic interface between grid and electrolyzer stack. Section III reviews and compares the different power converter topologies and control schemes that satisfy the above necessities. In Section IV, electromagnetic transient (EMT) simulation results using the power system emulator from Typhoon Hardware-in-Loop (HIL) Inc. are provided to support the analysis. The paper is concluded in Section V.

## II. Power Electronics Interfaces for Electrolyzers

This section discusses the equivalent model of the electrolyzer stack and the base requirements for a power electronics interface that allow the electrolyzer system to serve grid services. In this paper, we only focus on the PEM

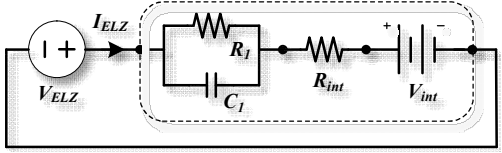

Figure 1. Equivalent model of a PEM electrolyzer.

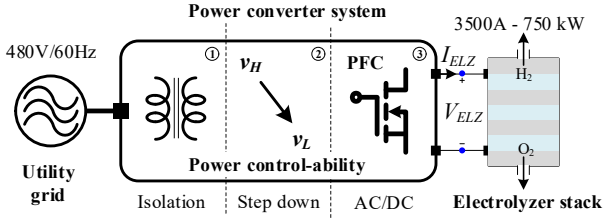

Figure 2. The base requirements for the power electronics interface between grid and electrolyzer stack (a representative configuration).

electrolyzers, since its fast response is suitable for grid services.

A. *PEM Electrolyzer Modeling*

The PEM electrolyzer can be model as an electrical circuit shown in Fig. 1 [13], where the temperature and anode and cathode pressures of the electrolyzer stack are assumed to be controlled at an appropriate level through low-level controls of the electrolyzer system (including stack, balance-of-plant, and hydrogen storage). The circuit parameters represent the reversible voltage ($V_{int}$) of the cathode, the membrane resistance ($R_{int}$), the double-layer capacitive effect ($C_1$), and the activation losses of the cathode $R_1$). For the stack of 3 cells, these parameters are given values of $V_{int}$ =4.38 V, $R_{int}$ =0.088 Ω, $R_1$ =0.035 Ω and $C_1$ =37.26 F with the operating current range from 0-50 A [14]. Considering only the base control purposes at the steady-state, the equivalent circuit can be reduced to the reversible voltage ($V_{int}$) and the internal resistance ($R_{total}$).

For the electrolyzer stack of 3500 A–750 kW, we could use the stack of 70 modules in parallel with 35 cells in series for one module. The equivalent reversible voltage for this stack can be established at 145.5 V and the total resistance is 0.02 Ω. Therefore, for the steady-state operation without considering the start-up process, when the electrolyzer current of the stack ranges from 0 to 3550 A, the stack voltage would change from 145.5 V to 215.5 V.

B. *Power Electronics Interface Baseline Requirements*

For a sake of unity, we consider the power converter system interfacing the utility grid at the line-to-line RMS value of 480 VAC and supplying it to the 3500 A-750 kW electrolyzer stack. A representative configuration used in this paper with the baseline requirements of this power electronics interface is depicted in Fig. 2. It should be able to accomplish three main conversion functions as 1) *galvanic isolation for protecting the electrolyzer stack from the stray currents*; 2) *voltage step-down to serve the electrolyzer at the rated DC voltage*; 3) *AC/DC conversion for providing DC power to the electrolyzer stack*. Additionally, the power conversion should not adversely impact the grid power quality with high harmonic content and be able to correct the power factor following the grid codes [14].

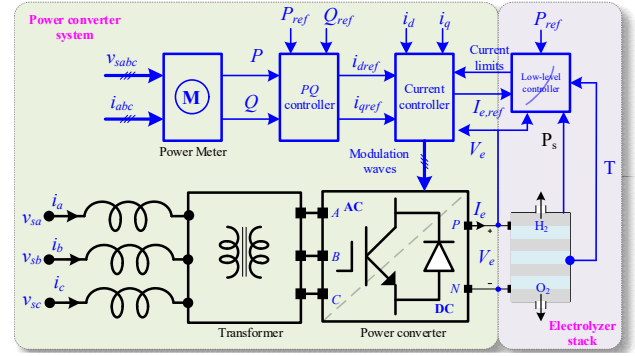

Figure 3. Control diagram of the electrolyzer converter system.

For the control requirements, the power converter system should be able to control both active and reactive power at two power operation quadrants, i.e., the electrolyzer system can consume active power but can absorb and inject reactive power [15]. To ensure the safety and physical integrity of the electrolyzer system, the power converter should work in coordination with the low-level controls for the electrolyzer system as well as follow grid interconnection and codes from the utility grid as shown in Fig. 3. The reference active and reactive powers ($P_{ref}$ and $Q_{ref}$) are derived from the high-level control which optimally regulates the operation of the electrolyzer system according to the grid-side conditions. The power converter controller will control the reference power while maintaining acceptable power quality for both the grid-side and electrolyzer-side by regulating the grid-side currents. The electrolyzer stack controller coordinates with the power converter system via $P_{ref}$ at its operating conditions of temperature (T) and pressure ($P_s$). The electrolyzer stack can also provide feedback for its current limits to the power converter system.

From the system point of view, the overall control architecture is depicted in Fig. 4. The power converter would host both the high-level controller for grid services, which includes the system control functions; and the power converter controller, which executes the application control functions and converter control. The system control functions can include the dispatch functions, autonomous response following frequency-watt, volt-var or volt-watt, etc. [15].

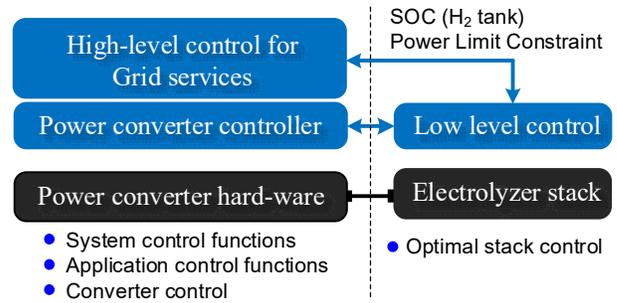

Figure 4. Control architecture of electrolyzer system for grid services.

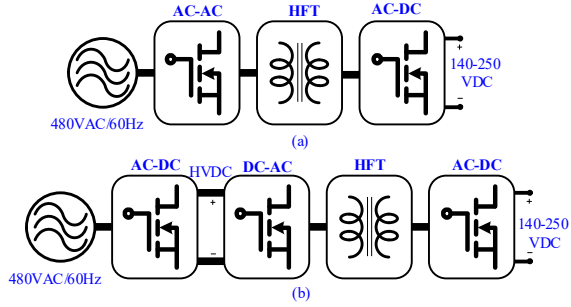

*Figure 5. SST topologies with (a) single-stage isolated AC-DC converters and (b) two-stage system of active rectifier and DAB converter.*

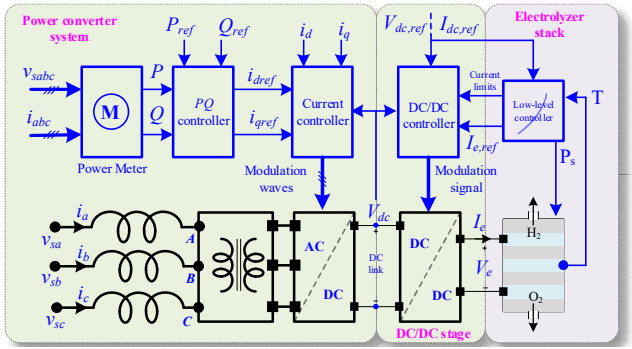

*Figure 6. Control diagram of the electrolyzer converter system with DC/DC stage.*

The application control takes care of voltage or frequency ride-through functions, whereas the converter control includes the phase lock loop synchronization, current control, switching control, and hardware control. The electrolyzer stack can also be controlled optimally while informing its state of charge (SOC) of the $H_2$ tank, allowable current limit, and ramp rate limit constraints to the high-level control.

### III. COMPARISON OF POWER CONVERTER TOPOLOGIES

The major consideration which would shape the design of such a power converter system for electrolyzer applications is the *galvanic isolation*. One should decide on using a grid-frequency transformer or a high-frequency transformer [16]. Some other issues which may also be considered are *power decoupling circuits*, *integration of ultra-capacitor* with various variations [17].

#### A. Solid-state Transformer (SST)

Due to the high frequency, the SST can significantly reduce its size and weight compared to the low-frequency transformer. In addition, SST provides more functionalities of actively regulate voltage and current, power flow control, and power quality regulation. Four possible SST configurations convert from high voltage AC to low voltage AC [18]. However, in electrolyzer applications, the conversion from high voltage AC to low voltage DC is required, so there are two possible configurations: 1) single-stage using isolated AC-DC converters, 2) two-stage using AC-DC converters, and then isolated DC-DC converters.

The single-stage SST configuration shown in Fig. 5 (a) is derived from [19], where an isolated AC-DC boost or the AC-DC dual active bridge (DAB) can be used to fulfill three main conversion functions for electrolyzer applications. In Fig. 5 (b), the two-stage SST configuration based on AC-DC active rectifier and DAB converter is described. These configurations are interfacing to a single-phase grid. Then, for the three-phase grid, one more switching leg should be added.

The concept of SST is interesting and potentially enables seamless grid integration; however, they are not mature yet, especially their control functionalities. In this paper, we only review their topologies and will be evaluated in detail in the future.

#### B. Grid-frequency Transformer

The traditional grid-frequency transformer has the weaknesses of voltage drop and losses at no load, sensitivity to harmonics and DC offset load imbalances. It also has a larger weight and volume compared to SST. However, it is relatively inexpensive, highly robust, and reliable, highly efficient, so it still dominates most power converter systems. When using a low-frequency transformer for isolation and step-down purposes, an active front-end rectifier is employed to regulate the consuming grid power. The DC/DC stage is optional for electrolyzer applications. Although this additional stage would reduce the conversion efficiency, it can provide benefits as follows:

- Increasing the controllable voltage range at the output. This increases the flexibility of designing a step-down transformer and enables the operation of an active rectifier at constant DC-link.

- Decoupling the AC input and DC output powers. This reduces the DC capacitor size.

- Separating the AC-side control and DC-side control.

The control diagram of the power converter system without DC/DC stage is the same as Fig. 3, where the power converter is only an AC-DC active front-end rectifier, which must regulate the AC-side and take care of the DC-side at the same time. This would limit the system's operation and reduce flexibility. Adding the DC/DC stage, the active rectifier can only regulate the active and reactive powers at the grid side, whereas the DC/DC converter can control the DC-side current voltage as shown in Fig. 6. The DC-link voltage can be regulated as a constant level to stabilize the power regulation and increase the power quality at the AC side.

More details of this comparison will be provided with simulation results in Section IV, where the power converter system with and without DC/DC stage is simulated with Typhoon HIL.

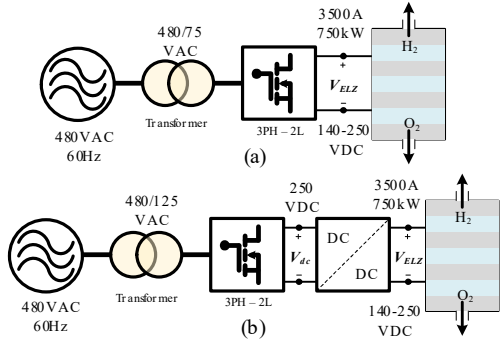

*Figure 7. Power converter system for electrolyzer applications (a) without, and (b) with DC/DC converter.*

## C. Other Configurations

One may want to use a buffer circuit to decouple the AC and DC power when working with a single-phase grid, then the size of the electrolytic capacitor can be reduced significantly or can be replaced by film capacitors [17]. However, it is not quite necessary at a three-phase grid. The integration of ultra-capacitor and its accompanied circuits can also be considered to ease the intermittent power allowing the electrolyzer stack operates at its optimal point [19].

## IV. DIGITAL SIMULATIONS USING TYPHOON HIL

In this section, we compare the power converter system with and without DC/DC stage. The detailed configurations of the two systems are given in Fig. 7, where the secondary voltage of the step-down transformer should design at a value that makes the DC voltage smaller than the designated value of the DC link or the electrolyzer reversible voltage when *controlling at the maximum of modulation waves*. In this paper, the DC voltage is chosen as 250V and 140V respectively; therefore, the voltage ratio of the step-down transformer would be 480/125 and 480/75, respectively. The electrolyzer stack has a rated current and power of 3500A–750kW with modeling parameters described in section II.A. Power system modeling details are presented in Appendix-A. The simulation results, which all are of 50 *ms*, for these two configurations using Typhoon HIL are provided as follows.

## A. Without DC/DC Stage

When there is no DC/DC stage, the active rectifier will uncontrollably charge the electrolyzer stack close to its reversible voltage level. Thanks to the intrinsic characteristic of PEM electrolyzer, the electrolyzer current is relatively small at this stage. The consumed power is mostly used to charge the PEM stack rather than producing hydrogen. Fig. 8 shows the operation of the electrolyzer system at a beginning state when the stack voltage is just a little bit higher than the reversible voltage. The consumed power is controlled at only 2 kW. The input currents are distorted due to their relatively small value of 13.45 A. The DC voltage of the stack is at 145.78 V. The switching voltages ($V_{ab}$, $V_{bc}$, and $V_{ca}$) of the three-phase two-level are shown at the bottom scope. When the active and reactive powers are controlled at 200kW and 50kVAr, respectively, the input current is smooth with nearly no harmonics as shown in Fig. 9. The electrolyzer stack current and voltage are 1155.96 A and 168.48 V, respectively, making the consumed power of the electrolyzer stack at about 194.756 kW with the conversion efficiency of 97.378 %. Then, the reference active power is increased to 500 kW. The transient response spans around 3ms. The steady-state values of DC current and voltage are 2501.54 A and 194.88 V, respectively, resulting in 487.5 kW of power consumption with 97.5% of the AC/DC conversion efficiency.

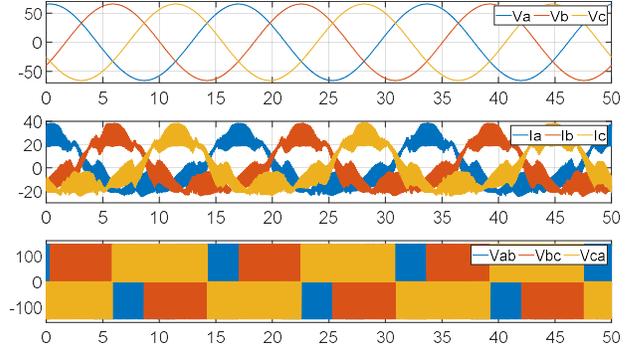

*Figure 8. Power converter system without DC/DC stage operating at the reference active power of 2 kW.*

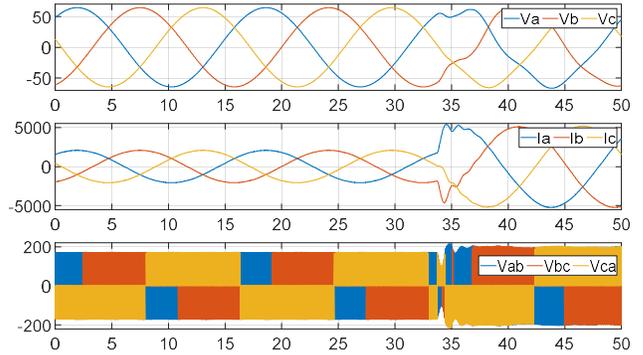

*Figure 9. Power converter system without DC/DC stage operating at the reference active and reactive powers of 200 kW-500 kW and 50 kVAr.*

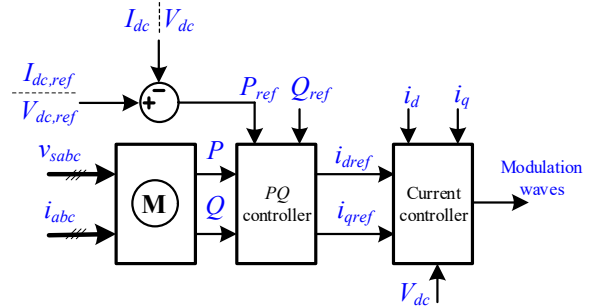

*Figure 10. $V_{dc}$-Q, $I_{dc}$-Q control loops for electrolyzer application*

The simulation results show that, even without DC/DC stage, the power converter system can control the electrolyzer stack nearly from 0–100% of its rated current. However, it cannot control accurately the grid-side powers and the DC-side current at the same time. *The DC-side current and voltage values depend on the DC power and the equivalent resistance of the electrolyzer stack.* The DC power equals the difference of the grid-side active power and losses. Instead of using the *P-Q* control loop, the $V_{dc}$-Q or $I_{dc}$-Q control loops as shown in Fig. 10 can directly regulate the DC-side voltage or current values, sequentially, *the active power depends on the DC-side voltage or current and the equivalent resistance of the electrolyzer stack*. Notably, the DC-side voltage and current cannot be controlled lower than its minimum values due to the boost feature of the active rectifier. Due to the page limit, the simulation results of these two control loops are not shown.

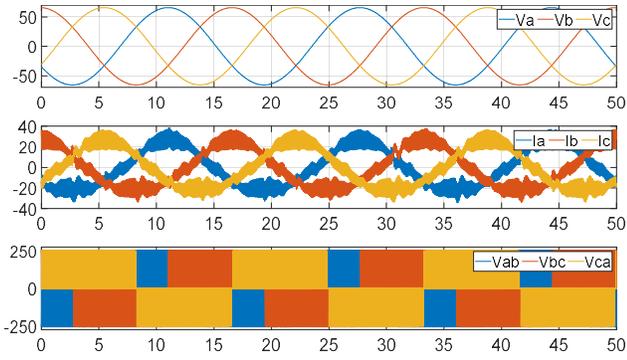

*Figure 11. Power converter system with DC/DC stage operating at the DC-link voltage of 250 V and the active power of 2 kW.*

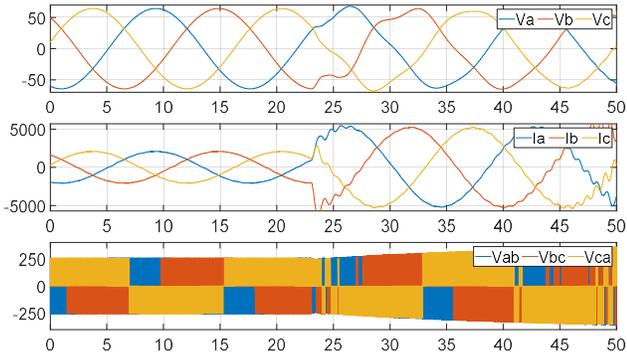

*Figure 12. Power converter system with DC/DC stage operating at the DC-link voltage of 250 V and the power of 200-500 kW and 50 kVAr.*

### B. With DC/DC stage

Using DC/DC stage helps to maintain the DC-link voltage at a constant value. Fig. 11 shows the operation with the DC-link voltage of 250 V and the active power of 2 kW. Compared to Fig. 8, the grid-side currents are less distorted due to high DC-link voltage. The electrolyzer voltage and current are 145.84 V and 13.03 A, respectively, making the 1.9 kW DC power with the conversion efficiency of 95 %. Fig. 12 shows the transient response when increasing the active power from 200 kW to 500 kW. The DC-link voltage is still controlled at 250 VDC. The electrolyzer voltage and current change from 168.39 V and 1140.2 A to 194.75 V and 2464.7 A, respectively. The conversion efficiency is 96%. The transient response is similar to the case without DC/DC stage. Adding DC/DC stage enables the operation at a certain level of the DC-link voltage. It can also provide better start-up progress of the electrolyzer system with a less distorted input current. However, it will reduce the total efficiency of the system and costs more for its power and control circuits. Considering the overall benefits, one may omit this DC/DC stage to reduce cost and increase efficiency.

### C. Electrolyzer Supporting Electric Grid under Load Change

In this case, a scaled real-time simulation of the electric grid with the detailed electrolyzer converter system without DC/DC stage is performed as shown in Fig. 13. The synchronous engine-generator source operates at 480V-60Hz with 3 MVA rated power, where the first-order generator model and speed proportional-integral (PI) regulator is used to simulate this generator in Typhoon HIL. This generator is supported by the 3500A-750 kW electrolyzer converter system through the 480/64 voltage ratio and 60 Hz transformer. The dynamic load has a nominal voltage of 480V and nominal power of 1 MVA. Simple rule-based control is used to maintain a constant net power flow at the generation bus. As shown in Fig. 14, there is a 400-kW load change in the dynamic load, the sub-second response from the electrolyzer is triggered to compensate for the sudden change. The generation power $P_G$ re-attains a steady-state at 1150 kW with a sub-second response from the electrolyzer. It may be noted that the electrolyzer response includes a delay representative of the telemetry. A more accurate characterization will be done with controller- and power-hardware-in-loop with electrolyzer hardware at NREL. When the load change occurs, the frequency deviates from 60 Hz, but under the response of the electrolyzer system, the frequency can be revived to 60 Hz. Therefore, the electrolyzer system can support the electric grid in maintaining power flow and frequency.

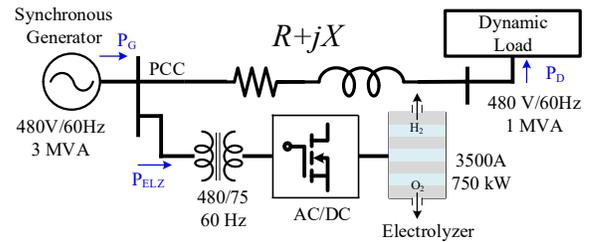

*Figure 13. A scaled electric grid simulation with dynamic load and electrolyzer system.*

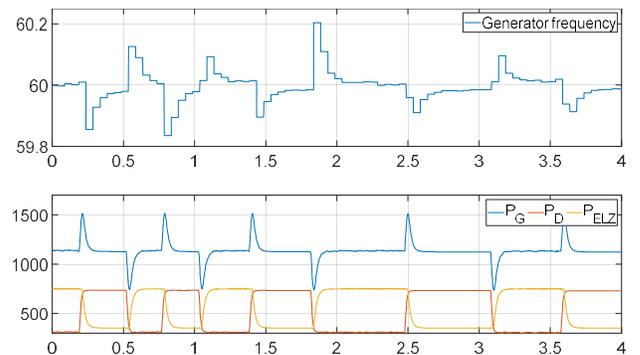

*Figure 14. Dynamic load changes and the response of electrolyzer system for compensation.*

## V. CONCLUSION

The paper provides a review on power converter topologies that have the ability of optimal power control for electrolyzers as dynamic loads, thereby enabling the grid services. As analyzed, the solid-state transformer could provide small footprints and better efficiency than the grid-frequency transformers. However, due to cost and reliability challenges, a traditional transformer is used for near-term integration. A more detailed analysis using SST will be presented in future work. The paper also provides the comparison between using DC/DC stage and without it. Although DC/DC stage enhances the control ability and enables the operation at constant DC-link voltage, it impacts the overall energy conversion efficiency of the system. Owing to the reversible voltage in the electrolyzer stack model, and the size of the stack, the electrolyzer can have a dynamic operation range (e.g., 145.5 V for 750kW stack used in this paper). With larger stack sizes, the stack configuration, and degradation of stack operational voltages the dynamic operational range can change. In such cases, power converter configuration with DC/DC stage can be beneficial and provide tighter regulation and improved power quality. For smaller stack sizes where high dynamic operation is not necessary, a single rectification stage without DC/DC stage may be more economical. In this paper, the digital real-time simulation results including the case of electrolyzer supporting electric grid under load change using Typhoon HIL have validated the performance of the two power converter topologies.

## ACKNOWLEDGMENT

This work was authored in part by the National Renewable Energy Laboratory, operated by Alliance for Sustainable Energy, LLC, for the U.S. Department of Energy (DOE) under Contract No. DE-AC36-08GO28308. Funding provided by U.S. Department of Energy Office of Energy Efficiency and Renewable Energy Hydrogen and Fuel Cell Technologies Office. The views expressed in the article do not necessarily represent the views of the DOE or the U.S. Government. The U.S. Government retains and the publisher, by accepting the article for publication, acknowledges that the U.S. Government retains a nonexclusive, paid-up, irrevocable, worldwide license to publish or reproduce the published form of this work or allow others to do so, for U.S. Government purposes.

## APPENDIX-A

*Electrolyzer power converter* ($R_{ac} = 10$ µΩ, $L_{ac} = 15.8$ µH, $f_{sw} = 10\ kHz$, Switch forward voltage drop $V_t = 1.7$ V, Diode forward voltage drop $V_t = 1.5$ V, $C_{dc1} = 7.5$ mF); *DC/DC stage* ($R_{dc} = 0.1$ mΩ, $L_{dc} = 50$ mH, $C_{dc2} = 1.2$ mH, same switching frequency and switch model); *Line impedance* ($R = 4$ mΩ, L = 44 µH); *Dynamic load* (power ref. change rate 100 p.u./s); *Generator* (first-order model with speed PI regulator $K_p = 20$, $K_i = 10$, $L_s = 50$ µF, $R_s = 1$ mΩ, the moment of inertia $J_g = 35$ kg.$m^2$).